\documentclass{jetpl}
\twocolumn

\usepackage{graphicx}% Include figure files

\lat

\title{Role of electronic correlations in the Fermi surface formation of Na$_x$CoO$_2$}

\rtitle{Electronic correlations in Na$_x$CoO$_2$}

\sodtitle{Role of electronic correlations in the Fermi surface formation of Na$_x$CoO$_2$}

\author{A.\,Shorikov $^+$, M.\,M.\,Korshunov $^{*,\dag}$\/\thanks{e-mail: korshunov@phys.ufl.edu, present address: Department of Physics, University of Florida, Gainesville, Florida 32611, USA}, V.\,I.\,Anisimov $^+$}

\rauthor{A.\,Shorikov, M.\,M.\,Korshunov, V.\,I.\,Anisimov}

\sodauthor{Shorikov, Korshunov, Anisimov}

\address{$^+$ Institute of Metal Physics, Russian Academy of Sciences-Ural Division, 620219 Yekaterinburg GSP-170, Russia\\
$^*$ L.V. Kirensky Institute of Physics, Siberian Branch of RAS, 660036 Krasnoyarsk, Russia\\
$^\dag$ Max-Planck-Institut f\"{u}r Physik komplexer Systeme, D-01187 Dresden, Germany}

\abstract{Band structure of metallic sodium cobaltate Na$_x$CoO$_2$ ($x$=0.33, 0.48, 0.61 0.72) has been investigated by local density approximation+Hubbard $U$ (LDA+$U$) method and within Gutzwiller approximation for the Co-$t_{2g}$ manifold. Correlation effects being taken into account results in suppression of the $e'_g$ hole pockets at the Fermi surface in agreement with recent
angle-resolved photo-emission spectroscopy (ARPES) experiments. In the Gutzwiller approximation the bilayer splitting is significantly reduced due to the correlation effects. The formation of high spin (HS) state in Co $d$-shell was shown to be very improbable.}

\PACS{74.70.-b, 31.15.Ar, 71.45.Gm, 71.10.-w}

\begin{document}

\maketitle

\textbf{1. Introduction.} Puzzling properties of sodium cobaltate Na$_x$CoO$_2$
are the topic of many recent theoretical and experimental investigations ~\cite{MMKreview}. This material holds much promise for thermoelectronics due to its large thermopower~\cite{thermopower} together with the relatively low
resistivity~\cite{Takada03}. The discovery of superconductivity with T$_c$
about 5K in Na$_{0.33}$CoO$_2$$\cdot1.3$H$_2$O~\cite{Takada03} revived the
interest in lamellar sodium cobaltates. Moreover, the charge and magnetic long
range orders on the frustrated triangular lattice of cobaltate is of the
fundamental interest. The band theory predict the complicated Fermi surface
(FS) with one large hole pocket around the $\Gamma=(0,0,0)$ point and six small
pockets near the ${\rm K}=(0,\frac{4\pi}{3},0)$ points of the hexagonal
Brillouin zone at least for $x<0.5$~\cite{Singh00,Pickett04}. However,
intensive investigations by several ARPES groups reveal absence of six small
pockets in both Na$_x$CoO$_2$$\cdot y$H$_2$O and in its parent compound
Na$_x$CoO$_2$ \cite{Hasan04,Yang04,Yang05,Qian06,Qian06_2}.

The disagreement between ARPES spectra and {\it ab-initio} calculated band
structure points to the importance of the electronic correlations in these
oxides. Other evidences for the correlated behavior come from the data on an
anomalous Hall effect and a drop of the thermopower in holistic magnetic
field~\cite{fluctuation}.

The six hole pockets are absent in the L(S)DA+$U$
calculations~\cite{Zhang04,Zou04}. However, in this approach, the insulating
gap is formed by a splitting of the local single-electron states due to
spin-polarization, resulting in a spin polarized Fermi surface with an area
twice as large as that observed through ARPES. Moreover, the long range
ferromagnetic order has been set by hand because of limitation of LDA+$U$. The
predicted large local magnetic moments as well as the splitting of bands can be
considered as artifacts of the L(S)DA+$U$ method.

Although LDA+$U$ method is usually applied to describe insulators \cite{Anisimov91}, there are some achievement in investigation of metals and metallic
compounds~\cite{Kotliar01,Mazin03}. To analyze the effect of electronic
correlations on the Fermi surface formation in sodium cobaltate we employ
non-magnetic LDA+$U$ method. Then, we use a Gutzwiller approximation to display the effect of correlations on the bilayer splitting and compare it with LDA+$U$ results.

Co $d$-level splits by crystal field of oxygen octahedron in lower $t_{2g}$ and higher $e_g$ bands. The deficiency of Na in Na$_x$CoO$_2$ introduces additional
holes in the system. Cobalt, having $d^6$ configuration and filled $t_{2g}$
shell in parent NaCoO$_2$, is nonmagnetic. But in nonstoichiometric compound
part of Co ions become magnetic with local moment about 1$\mu_B$. This value is provided by $d^5$ configuration and one hole in $t_{2g}$ shell. However, the
experiments revealed the magnetic susceptibility at room temperature that is
much higher than it was expected for dilute magnetic impurity in non-magnetic
solvent. Explanation of this anomaly was suggested in Ref.~\cite{daghofer06} as
transition from low-spin (LS) state with six $d$-electrons on $t_{2g}$ shell to high-spin (HS) state with five $d$-electrons on $t_{2g}$ shell and one electron on $e_g$. The possibility of such transition will be discussed below.

\textbf{2. LDA+$U$ results.} Na$_{0.61}$CoO$_2$ crystallize in the hexagonal
unit cell ($P6_{3}/mmc$ space group) with $a$=2.83176(3)\AA~  and $c$=10.8431(2)\AA~ at 12K Ref.~\cite{Jorgensen03}. Displacement of Na atoms
from their ideal sites $2d$ $(1/3,2/3,3/4)$ on about 0.2\AA~ are observed in
non-stoichiometric cobaltates for both room and low temperatures. This is
probably due to the repulsion of a randomly distributed Na atoms, locally
violating hexagonal symmetry~\cite{Jorgensen03}. In the present investigation
Na atoms are shifted back to their 2d ideal sites. In order to avoid charge
disproportionately which can arises from some Na distribution if the supercell
is used in calculation, the change in the Na concentration has been considered
in virtual crystal approximation (VCA) where each 2d site is occupied by virtual atoms  with fractional number of valence electrons $x$ and a core charge $10+x$ instead of Na. Note, that all core states of virtual atom are left unchanged and corresponds to Na ones. We have chosen $4s$, $4p$, and $3d$ states of cobalt, $2s$, $2p$, and $3d$ states of oxygen, and $3s$, $3p$, and $3d$ states of Na as the valence states for TB-LMTO-ASA computation scheme. The radii of atomic spheres where 1.99 a.u. for Co, 1.61 a.u. for oxygen, and 2.68 a.u. for Na. Two classes of empty spheres (pseudo-atoms without core states) were added to fill the unit cell volume.

Crystal field of oxygen octahedron splits Co $d$-band into doubly degenerate
$e_g$ and triply degenerate $t_{2g}$ subbands (without taking spin into account). LDA calculations shows that those manifolds are separated by about 2
eV~\cite{Singh00}. Here partially filled $t_{2g}$ subband crosses the Fermi
level whereas $e_g$ subband due to strongly hybridization with nearest oxygen
atoms is positioned well above the Fermi level. The procedure proposed in
Ref.~\cite{Gunnarsson} allows one to calculate the Coulomb repulsion parameter $U$ taking into account the screening of localized $d$-shell by itinerant $s$- and $p$-electrons. Resulting $U$ is equal to 6.7 eV. However, the presence of the $t_{2g}$--$e_g$ splitting give the reason to take into account an additional screening channel provided by the less localized $e_g$ electrons. The value of $U$=2.67 eV for $t_{2g}$ orbitals was calculated using the ``constrained LDA'' method~\cite{Pickett98}, where the screening by the $e_g$ electrons is also taken into account. This value was used in the
present calculation for all doping concentrations $x$. Hund's exchange
parameter $J$ depends weakly on screening effects due to its ``on-site''
character. Its value was also calculated within the ``constrained LDA'' method and was found to be 1.07 eV.

First, we have verified the possibility of HS state formation on Co $d$-shell.
For this purpose the unit cell of Na$_{0.61}$CoO$_2$ with two Co atoms was
considered. We have started from a saturated A-type antiferromagnetic
configuration with five electrons on the $t_{2g}$ and one on the $e_g$ shells.
Small $U$=2.67 eV does not stabilize such magnetic configuration and LS state
was obtained. Increasing $U$ up to 5 eV however results in HS state with large
local magnetic moment about 1.96 $\mu_B$. Nevertheless, this HS state has the
total energy about 1.75 eV higher then the energy of a LS state. This large
difference in total energy of both considered spin states arises form the hexagonal structure of cobaltates where the angle of Co-O-Co bond is close to 90$^\circ$ in contrast to almost 180$^\circ$ in, e.g., RCoO$_3$ (R=La, Ho). In the latter case the $e_g$ band has the width of about 3-5 eV and its bottom lies just above the Fermi level. The system wins energy of 2$J$ forming a HS state overcoming the gap energy which is less than 1 eV. Due to this fact the difference between the LS and intermediate spin states in RCoO$_3$ is less then 250 meV~\cite{Nekrasov03}. The angle of Co-O-Co bond is close to 90$^\circ$ in
cobaltates and it results in a weak overlap between $e_g$ orbitals and hence in a narrow $e_g$ band with larger gap between it and the $t_{2g}$ band. Our
calculation confirms that formation of the HS state in Na$_x$CoO$_2$ is rather improbable and cannot be stabilized by any distortion of crystal structure or
clusterization proposed in Ref.~\cite{daghofer06}. Local magnetic moments on Co sites can arise only because of the doped holes due to Na atoms deficiency. Those holes order on Co atoms and form nonmagnetic Co$^{3+}$ and magnetic Co$^{4+}$ ions with $d^6$ and $d^5$ configurations, respectively. In the following, we consider only the LS state.

The ordering of holes on $t_{2g}$ shell and corresponding long-range magnetic
and charge orders in Na$_{0.5}$CoO$_2$ arise probably due to specific
arrangement of Na atoms. These arrangements were observed
experimentally~\cite{ordering} for several doping concentrations including
$x=0.5$. Proper description of such order within the ``unrestricted Hartree-Fock'' gives strong spin and orbital polarization and local magnetic moment of about 1$\mu_B$ on Co$^{4+}$ sites as well as the insulating ground state with a sizable gap. To describe the non-ordered systems, the implementation of the ``restricted Hartree-Fock'' method is more suitable. In the latter, starting from the non-magnetic configuration of the $d$-shell with the equal number of spin-up and spin-down electrons, LDA+$U$ method gives the non-magnetic solution without spin or orbital polarization. Note, that the gap does not open and Na$_x$CoO$_2$ remains metallic for all Na concentration.

\begin{figure}
\centering
\includegraphics[width=0.8\linewidth]{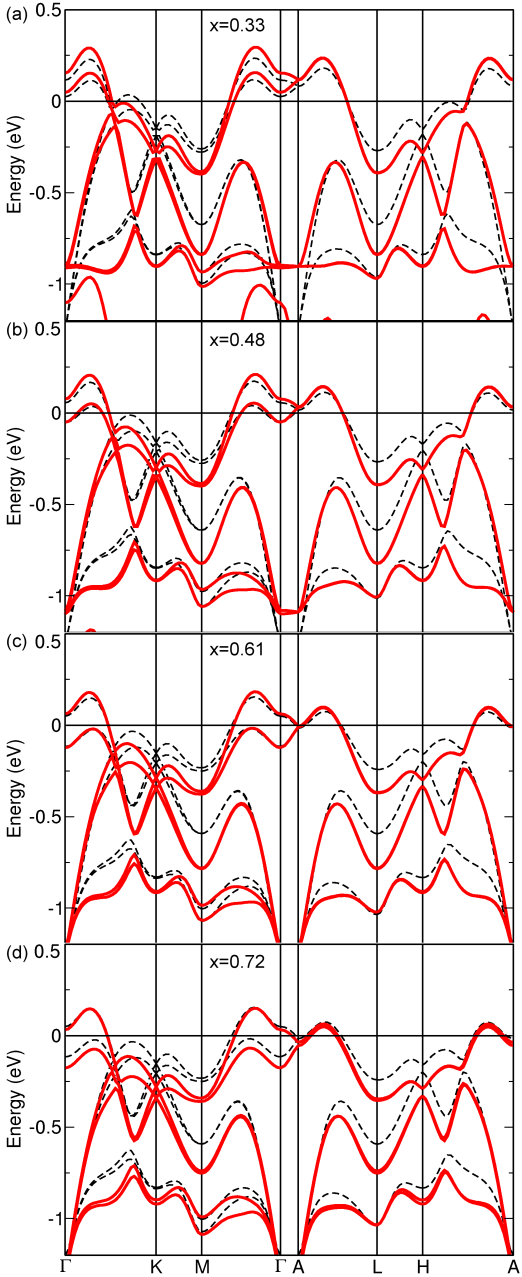}
\caption{\textbf{Fig. 1.} Band structure of Na$_x$CoO$_2$ for $x$ equal to 0.33 (a), 0.48 (b), 0.61 (c), and 0.72 (d), obtained within LDA is shown by the black (dashed) curves. Band structures for the same doping concentrations within LDA+U are shown by the red (solid) curves.}
\label{fig:bnd}
\end{figure}

Obtained band structure of Na$_x$CoO$_2$ for $x$=0.33, 0.48, 0.61, and 0.72 are shown in Fig.~\ref{fig:bnd}. Dashed (black) lines correspond to LDA results whereas solid (red) lines are the bands obtained by LDA+$U$ method. Cobalt $d$ and oxygen
$p$ states are separated by a small gap of about -1.25 eV for $x=0.61$ and
$x=0.72$. However, this gap disappears for lower doping concentration since the $d$ band goes down when the number of $d$ electrons decreases. The presence of
the two CoO$_2$ layers within the unit cell due to alternation of the oxygen
arrangement results in a bonding-antibonding (bilayer) splitting, also present
in Fig.~\ref{fig:bnd}.

The degeneracy of the $t_{2g}$ levels is partially lifted by the trigonal
crystal field distortion which splits them into the higher lying $a_{1g}$ singlet and the two lower lying $e'_g$ states. However, slight difference in
occupation numbers of $a_{1g}$ and $e'_g$ orbitals (0.714 and 0.886
respectively for $x=0.33$) results in a significant difference between the LDA+$U$ and LDA band structures. The energy of the less occupied $a_{1g}$
orbital increases for both spins, whereas all $e'_g$ bands go down (the total
$a_{1g}$--$e'_g$ splitting becomes 0.21 eV for $x=0.33$). This makes six $e'_g$ Fermi surface hole pockets to disappear for small $x$ values. Note that for all Na concentration LDA+$U$ predicts large $a_{1g}$ Fermi pocket centered
around the $\Gamma$ point in excellent agreement with the ARPES spectra for
$x<0.7$. The additional electron pocket close to the $\Gamma$ point appears in
both LDA and LDA+$U$ methods for a large doping concentrations. It was
discussed in our previous work~\cite{Korshunov06,Korshunov07} in connection
with the electronic theory for the itinerant magnetism of highly doped compounds.

\textbf{2. Gutzwiller approximation.} For the small doping concentrations, $x
\approx 0.3$, sodium cobaltate displays a canonical Fermi-liquid behavior both
in resistivity \cite{Foo2004} and in NMR relaxation rate \cite{Ning2004}.
Transport measurements \cite{Li2004} on single crystals with $x=0.7$ also
revealed Fermi-liquid behavior at low temperatures. However, this behavior is
characterized by the enormous electron-electron scattering. The Gutzwiller
approximation \cite{gutzwiller1963,gebhard1990,kotliar1986} for the Hubbard model recommended itself as a good tool to describe low-energy quantities such as the FS and a ground state energy in the correlated metallic system. We will use the multiband generalization of this approximation \cite{gebhard1997} to
investigate the effect of correlations on the bilayer splitting and compare it
with the LDA+$U$ results.

Hamiltonian for CoO$_2$-plane in a hole representation is given by:
\begin{eqnarray}
H = &-& \sum\limits_{{\bf f},\alpha ,\sigma }
\varepsilon^\alpha n_{{\bf f} \alpha \sigma } -
\sum\limits_{{\bf f} \ne {\bf g}, \alpha, \beta, \sigma}
t_{{\bf f}{\bf g}}^{\alpha \beta } d_{{\bf f} \alpha \sigma }^\dag d_{{\bf g} \beta \sigma} \nonumber\\
&+& \sum\limits_{{\bf f}, \alpha} U_\alpha n_{{\bf f} \alpha \uparrow} n_{{\bf f} \alpha \downarrow},
\label{eq:HG}
\end{eqnarray}
where $d_{{\bf f} \alpha \sigma}$ ($d_{{\bf f} \alpha \sigma}^\dag$) is the
annihilation (creation) operator for the $t_{2g}$ hole at Co site ${\bf f}$,
spin $\sigma$ and orbital index $\alpha$, $n_{{\bf f} \alpha \sigma} = d_{{\bf
f} \alpha \sigma}^\dag d_{{\bf f} \alpha \sigma}$, and $t_{{\bf f}{\bf
g}}^{\alpha \beta }$ is the hopping matrix element between two lattice sites
connected by the spatial vector $({\bf f}-{\bf g})$, $\varepsilon^{\alpha}$ is
the single-electron energies in which the chemical potential $\mu$ is included. Since LDA-calculated hoppings and single-electron energies do not depend much
on doping concentration~\cite{Korshunov06,Korshunov07}, we use here parameters
for Na$_{0.33}$CoO$_2$ form Table~I of Ref.~\cite{Korshunov07}. To take the bilayer splitting into account, we also consider hoppings $t_z^{\alpha \beta}$ between the adjacent CoO$_2$ planes. Their values (in eV) were also derived from LDA
results and are equal to $t_z^{a_{1g} a_{1g}}=-0.0121$, $t_z^{e'_{g1}
e'_{g1}}=0.0080$, and $t_z^{e'_{g2} e'_{g2}}=-0.0086$.

\begin{figure}
\centering
\includegraphics[width=0.8\linewidth]{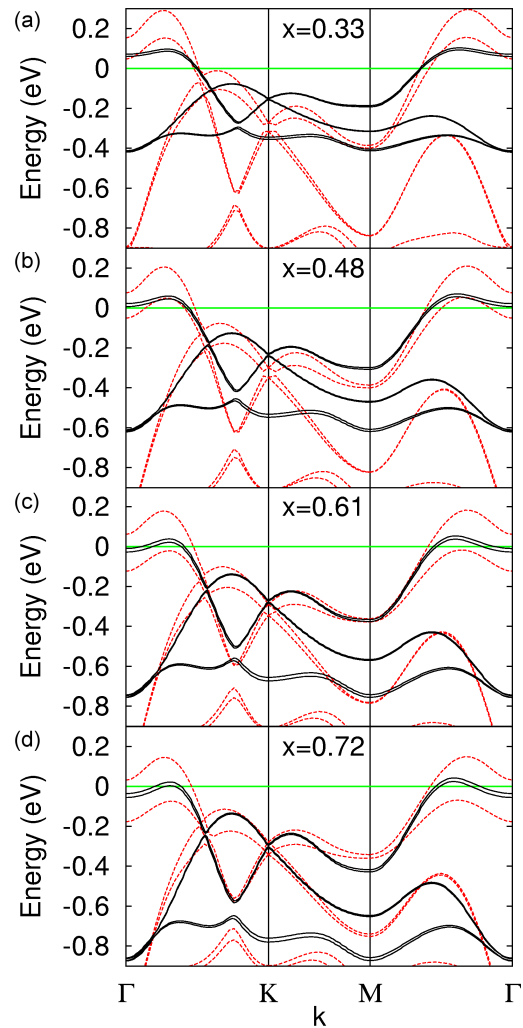}
\caption{\textbf{Fig. 2.} Band structure of Na$_x$CoO$_2$ for $x$ equal to 0.33 (a), 0.48 (b), 0.61 (c), and 0.72 (d), obtained in LDA+U is shown by the dashed (red) curves. Dispersion within the Gutzwiller approximation is shown by the solid (black) curves.}
\label{fig:gutzwiller}
\end{figure}

Within the Gutzwiller approximation the Hamiltonian describing the interacting
system far from the metal-insulator transition for $U \gg W$,
$J=0$, is replaced by the effective non-interacting Hamiltonian:
\begin{eqnarray}
H_{eff} = &-& \sum\limits_{{\bf f}, \alpha, \sigma}
\left( \varepsilon^{\alpha} + \delta\varepsilon^{\alpha\sigma} - \mu \right) n_{{\bf f} \alpha \sigma} \nonumber\\
&-& \sum\limits_{{\bf f} \ne {\bf g}, \sigma} \sum\limits_{\alpha, \beta}
{\tilde t_{{\bf f}{\bf g}}}^{\alpha \beta} d_{{\bf f} \alpha \sigma}^\dag  d_{{\bf g} \beta \sigma} +
\sum\limits_{\alpha, \sigma} \delta\varepsilon^{\alpha \sigma} n_{\alpha \sigma}.
\label{eq:Heff}
\end{eqnarray}
Here, ${\tilde t_{{\bf f}{\bf g}}}^{\alpha \beta} = t_{{\bf f}{\bf g}}^{\alpha
\beta} \sqrt{q_{\alpha \sigma}} \sqrt{q_{\beta \sigma}}$ is the renormalized
hopping, $q_{\alpha \sigma} = \frac{x}{1 - n_{\alpha \sigma}}$, $n_{\alpha
\sigma} = \left< n_{{\bf f} \alpha \sigma}\right>_0$ is the orbital's filling
factors, $x = 1 - \sum\limits_{\alpha \sigma} n_{\alpha \sigma}$ is the
equation for the chemical potential. $\delta\varepsilon^{\alpha\sigma}$ are the Lagrange multipliers yielding the correlation induced shifts of the
single-electron energies. They are determined by minimizing the energy
$\left<H_{eff}\right>_0$ with respect to the orbital filling factors $n_{\alpha \sigma}$. It is this energy shift $\delta\varepsilon^{\alpha \sigma}$, that
forces the $e'_g$ bands to sink below the Fermi level \cite{Zhou05}. This is
clearly seen in the doping-dependent evolution of the quasiparticle dispersion
within the Gutzwiller approximation in Fig.~\ref{fig:gutzwiller}. To obtain these figures we self-consistently solved equations on $\delta\varepsilon^{\alpha \sigma}$ and on the chemical potential $\mu$.

The comparison of the Gutzwiller approximation results with the LDA+$U$
dispersion reveals few very interesting conclusions. First, both approximations result in a suppression of $e'_g$ hole pockets of the FS. Second, the bilayer
splitting is strongly doping dependent and significantly reduced for the Gutzwiller quasiparticles in comparison with the LDA+$U$ quasiparticles because the renormalization coefficient, $\sqrt{q_{\alpha \sigma}} \sqrt{q_{\beta
\sigma}}$, occurs not only for the in-plane hoppings, but also for the
interlayer hoppings $t_z^{\alpha \beta}$. Third, when both bonding and
antibonding $t_{2g}$ bands do not cross the Fermi level around the $\Gamma$
point, the FS crossings are the same in both approximations (see Fig.~\ref{fig:gutzwiller}a). It is a simple consequence of the Luttinger theorem which holds for both approaches. But for large $x$ due to the larger bilayer splitting in the LDA+$U$ approach, the Fermi surfaces become different, while the Luttinger theorem is again preserved. With increase of the doping concentration $x$, the bandwidth of the Gutzwiller quasiparticles becomes closer to the LDA+$U$ because the band renormalization factor $\sqrt{q_{\alpha \sigma}} \sqrt{q_{\beta \sigma}}$ comes closer to unity.

Now we will discuss the correlations between our results and a more rigorous
theory, namely, the Dynamical Mean Field Theory (DMFT). Generally, within DMFT
the band structure of a Hubbard model consist of three parts: two incoherent
Hubbard subbands and a coherent near-Fermi-level quasiparticle band. Since $U$
is not very large in sodium cobaltates, Hubbard subbands loose their spectral
weight and are almost merged with the coherent band. Thus, low-energy
excitations are determined mostly by this quasiparticle band. And it is this
band that revealed within the Gutzwiller approximation, even if we formally in
the limit of $U \gg W$.

In the case of Na$_x$CoO$_2$, DMFT calculations show that for the small $U$ and a non-zero $J$, $e'_g$ FS pockets can be stabilized
\cite{Ishida2005,Perroni05}. On the other hand, more recent DMFT calculations of Ref.~\cite{Marianetti2006} confirms results of the Gutzwiller approximation
provided that the crystal field slitting $\Delta$ is about 50 meV. This
value is in agreement with our value of $\Delta=53$ meV \cite{Korshunov07}, so our results are consistent with DMFT.

\textbf{2. Conclusion.} In the present work employing {\it ab-initio} ``constrained LDA'' method we obtained Coulomb repulsion parameter $U$=2.67 eV
for $t_{2g}$ orbitals taking into account the screening by the $e_g$-electrons in addition to the screening by the itinerant $s$- and $p$-electrons. Hund's exchnage parameter was found to be $J$=1.07 eV.

Also we have shown that due to the Co-O-Co bond angle being close to 90$^\circ$ in Na$_x$CoO$_2$, the energy gap between the LS and HS states is too large to be overcome by the clusterization or reasonable distortions of the crystal structure. Thus we conclude that realization of the HS state is highly improbable in these particular substance.

To analyze the effect of electronic correlations on the Fermi surface
topology of Na$_x$CoO$_2$ we use two approaches, non-magnetic LDA+$U$ and
the Gutzwiller approximation for the Hubbard-type model based on the LDA band
structure. Physically, the reason of $e'_g$ FS pockets disappearance is quite clear. Within LDA+$U$ the energy of the less occupied $a_{1g}$ orbital
increases for both spins, whereas all $e'_g$ bands go down. This makes six
$e'_g$ FS hole pockets to disappear for small $x$ values, in agreement with
ARPES for $x<0.7$. Gutzwiller approximation also resulted in a suppression of
$e'_g$ hole pockets at the FS. Most importantly, the bilayer splitting was
found to be strongly doping dependent and significantly reduced for the Gutzwiller quasiparticles in comparison with the LDA+$U$ quasiparticles. This may explain why the bilayer splitting is not observed in ARPES though it is very pronounced in the LDA band structure.

Authors thank I. Eremin, M. Laad, and S.G. Ovchinnikov for useful discussions.
A.S. and V.I.A. acknowledge the financial support from RFBR (Project No. 10-02-00046-a, 09-02-00431-a and 10-02-00546-a), the fund of the President of
the Russian Federation for the support of scientific schools NSH 1941.2008.2,
the Programs of the Russian Academy of Science Presidium ``Quantum microphysics of condensed matter'' N7 and ''Strongly compressed materials``, Russian Federal Agency for Science and Innovations Project No. 02.740.11.0217, MK-3758.2010.2.
M.M.K. acknowledge support form INTAS (YS Grant 05-109-4891), RFBR (Grants
09-02-00127, 06-02-16100, 06-02-90537-BNTS), the Integration Program of SBRAS
N40, the Presidium of RAS Program 5.7, President of Russia (grant MK-1683.2010.2), FCP Scientific and Research-and-Educational Personnel of
Innovative Russia for 2009-2013 (GK P891).

\end{document}